# Development of Precise Low Value Capacitance Measurement System for Cryogenics Two Phase Flow Application

SINGH G. K.[1], PURWAR G., PATEL R., TANNA V.L., PRADHAN S.

[1] Institute for Plasma Research,
Homi Bhabha National Institute,
Bhat, Gandhinagar-382 428 Gujarat, India, E-Mail: gaurav.singh@ipr.res.in

*Abstract* – *In cryogenic two phase flow, it is always challenging to measure the quality and void fraction. In this regard, an effort has been made to indigenously develop an electronic circuit to measure the void fraction by measuring the capacitance of the order of picofarads accurately depending upon the dielectric constant of nitrogen in vapor and liquid phase. The state-of-art electronics card has been developed and tested successfully for its performance and validation. Using this card, an experiment has been conducted using liquid nitrogen cryo transfer line to study the two phase void fraction. In this paper, the design basis and working principle of the designed electronics card along with the performance results are discussed. A/D conversion and DAQ system has been implemented to display the direct measurement data in a computer.*

## I. INTRODUCTION

Two-phase flows are found in many areas of cryogenics including LNG, large-scale helium systems and space cryogenics [1]. Two phase flows are characterized by its void fraction. Void fraction is the most vital parameter for prediction of any hydrodynamic properties of fluid with two phase flow conditions. Void fraction of cryogen two phase flow could be measured by various techniques [2]. In this paper, we discuss innovative technique for void fraction measurement. It is based on capacitive void fraction measurement. The specific technique has been developed as it is simple to construct, low cost, non-intrusive and quick response. Capacitance of two phases measured with change in dielectric medium of cryogenic fluid. Here, we discuss the case of liquid nitrogen ($LN_2$) cryogen. Dielectric constant ($\varepsilon$) of nitrogen varies from 1.0 to 1.43 from vapor to liquid phase and the capacitance to be measured is in the range of picofarads (pF). There was a need of very precise and sensitive with high resolution electronics circuit that could measure capacitance in the range of picofarads. This work focuses on the development of an electronic circuit for the void sensor to measure the void fraction of the $LN_2$ two phase flows.

The design of an electronic circuit is an application of Schmitt trigger based IC CD4093B with capacitors in differential modes and trimmer capacitor for lead compensation. There are various challenging problems, which are encountered in measurement of very low capacitance measurement such as stray capacitance and lead capacitance. Apart from this, the drift and the offset errors, ambient temperature errors, range of frequency of operation, cost of development are also challenging in nature. In literature, various methods reported for capacitance measurement Viz. AC bridges, charge discharge methods, oscillation and resonance based methods[3-4], resonance based methods [5] etc. All above methods are accurate but not useful for picofarads capacitance measurement as it does not provide stray capacitance elimination. Resonance based method is somewhat better method but not useful for real time dynamic capacitance measurement. Presently, electrical capacitance tomography (ECT) method [6] uses integrator, which works on the principle of capacitor charge discharge method. However, an integrator induces drifts and offset errors, which are not desirable.

This present work provides a developed electronic circuit in which all above errors are taken into considerations to have precise and dynamic measurement of capacitance of the order of picofarads. The innovation of reported technique is that it uses charging and discharging method but it does not uses integrator so drift and offset errors are minimized. It uses quad two input NAND Schmitt trigger circuits CD4093B with hysteresis [7] and the electronic circuits have been successfully tested over a range of 0-600pF with percentage (%) error of $\pm\,0.58\%$. The present capacitance measurement work is inspired by the work reported by Gautam Sarkar et al, [8] and he measured low capacitance with lead capacitance compensation. Lead capacitance is major source of stray capacitance, which can be minimized by this method. However, system developed by Gautam [8] limited to 400 pF. The proposed work is an improved work in which the capacitance measurement range is increased with lead compensation up to 600 pF with linear input and output relationships. This built system is used for two phase liquid nitrogen void fraction measurement in terms of capacitance. The measured capacitance values were compared with highly sophisticated LCR digital meter. This system can also be used in real time dynamic capacitance measurements over a given range.

## II. SCHMITT-TRIGGER BASED VOID FRACTION AND CAPACITANCE MEASUREMENT SYSTEM

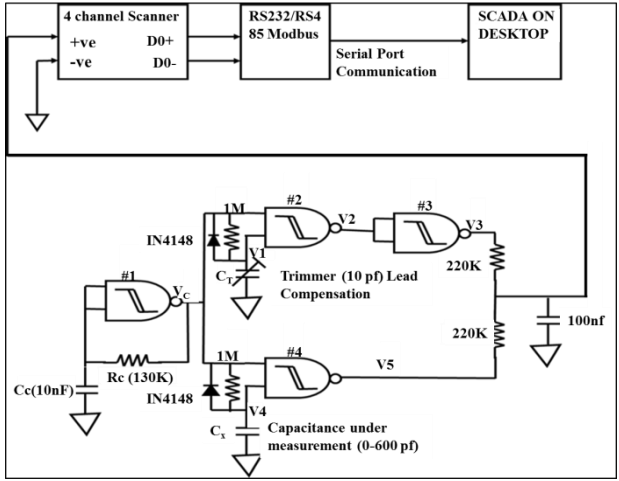

Fig. 1: Block diagram of the electronic circuit developed for capacitance measurement

Table 1: Required Hardware for electronic circuit development

| Sr. No. | Name of Component | Specifications | Required Quantity (Nos.) | Tolerance |
|---|---|---|---|---|
| 1. | Resistance | 130KΩ | 01 | ±5% |
| 2. | Resistance | 1MΩ | 02 | ±5% |
| 3. | Resistance | 220KΩ | 02 | ±5% |
| 4. | Capacitor | 10nF | 01 | ±1% |
| 5. | Capacitor | 100nF | 01 | ±1% |
| 6. | Trimmer Capacitor | 10pF | 01 | |
| 7. | Integrated Circuits | CD4093B Schmitt trigger | 01 | |
| 8. | Diode | IN4148 | 02 | |
| 9. | Power Supply | 5 Volts | 01 | |
| 10. | Printed Circuit Board | | 01 | |
| 11. | 3 pin PCB Power connector | | 03 | |
| 12. | Crocodile Clips | | 02 | |
| 13. | Connecting Wires | | ~01 meter | |

*A. Working Principle:*

Figure 1 shows the block diagram of the electronic circuit for capacitance measurement. The component used in electronic circuit development is listed in Table 1. In this circuit $C_T$ is the trimmer capacitor and $C_x$ is the capacitance that we want to measure. This circuit uses IC CD4093B which has 4 nos. of NAND gates. Each NAND gate of IC CD4093B having Schmitt trigger hysteresis property. Trimmer capacitor is used for lead capacitance compensation. Lead capacitance is a major source of stray capacitance. Capacitor of 100 nF is used for bypassing AC ripples into ground. The Schmitt trigger circuit having two threshold level such as upper threshold level and lower threshold level. When the input reaches beyond upper threshold value ($V_{UT}$) output goes to –Vsat and when the input reaches below lower threshold value ($V_{LT}$) then output reaches to +Vsat. In this way square wave generated. The band gap between upper threshold and lower threshold value is called hysteresis voltage ($V_H$).

$$V_H = V_{UT} - V_{LT}$$

In the figure 1 NAND gate 1 generates square wave of voltage ($V_C$) 5 volts with Schmitt trigger operation and time period of square wave is $T = 1.35\ ms$ duration and frequency of operation is 740.0 Hz. Since time period of the square wave is large so we get a wide range for capacitance measurement. In NAND gate 2, square wave of voltage $V_C$ and trimmer capacitor voltage $V_1$ is applied. Due to hysteresis property capacitor voltage $V_1$ becomes $V_{1H}$, Which creates time lag during the charging and discharging of capacitor. Time lag during discharging of capacitor is minimized by employing a diode. Thus diode reduces nonlinearity in the developed electronics circuit. Output voltage ($V_2$) of NAND gate 2 is feed to the common input of NAND gate 3. The output voltage of NAND gate 3 ($V_3$) is the inverted voltage of NAND gate 2 ($\overline{V}_2$). Time lag during charging and discharging is calculated by

$$T_1 = T_2 = RC_T ln(V/V_{TH}) = K_1 C_T$$

In NAND gate 4, square wave voltage $V_C$ and unknown capacitor voltage $V_4$ is applied. Due to hysteresis property of NAND gate unknown capacitor voltage $V_4$ becomes $V_{4H}$, Which creates time lag during charging and discharging of capacitor. Time lagging during discharging of capacitor is minimized by employing a diode and thus nonlinearity have been reduced in the electronics circuit. The output of NAND gate 5 is $V_5$. Time lag during charging and discharging is calculated by

$$T_3 = T_4 = RC_X ln(V/V_{TH}) = K_1 C_X$$

The output voltage $V_3$ and $V_5$ are applied to the potential divider circuit having 220K ohm resistance. Output voltage is tapped at the center of the voltage divider, which is filtered by the 100nF capacitor. 100 nF capacitor bypasses all AC particles to the ground and provides smooth DC voltage output $V_0$. This electronic circuit output voltage $V_0$ is linearly proportional to the difference of capacitances ($C_X$-$C_T$)

$$V_o = \frac{V}{2} + K_1(C_X - C_T)$$

Where $K_1$ = Proportionality constant.
Boundary conditions for the electronic circuit are
Voltage $V_3 = 0$ V and $V_5 = 5$ V then $V_0 = 2.5$ V.
And when $V_5 = 0$ V and $V_3 = 5$ V then $V_0 = 2.5$ V.
And when $V_3 = 5$ V and $V_5 = 5$ V then $V_0 = 5$ V.

Thus it is found that output voltage varies linearly in between of 2.5 V to 5 Volts.

*B. Performance test:*

The electronic circuit performance test is done after designing and development of capacitance measurement circuit. The electronics circuit is tested with several capacitors and operation frequency of 740 Hz.

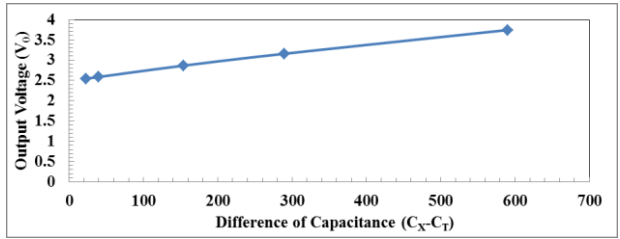

Fig. 2: Variation of output voltages with respect to difference of capacitances

From the figure 2 illustrate that the output voltage is directly proportional to the difference of unknown capacitance ($C_X$) and trimmer capacitance ($C_T$). There is a linear relation between output voltage of the developed electronic circuit and difference of the trimmer capacitance and unknown capacitance. Since range of the circuit is depends on the time period of the generated square wave, not on the output voltage ($V_o$). So the output voltage get saturated when capacitance value exceeded 600 pF value. In the performance test the range of capacitance measurement found in the range of 0-600 pF.

*C. Data acquisition system:*

Data acquisition and real time online monitoring of the process values are a challenging task. Data acquisition have been done with 4 channel 8204 masibus scanner, which provides front panel display of the process values. Scanner is connected with process analog input values at input channels and relay terminals. Channels are selected as per our process variable requirements. As per the requirement one analog channel is selected and necessary settings are done for communication with Modbus, desktop display, SCADA and Scanner.
The Output voltage $V_o$ of the electronic circuit is connected with the analog input channel of the scanner. The analog input channel number, upper, lower voltage range, baud rate, parity bit etc. is programmed manually. A/D conversion process is done in the scanner and it gives proportional count values for communication with SCADA. Output of the scanner (counts) is communicated to the PC via Modbus RS485 interface. RS485 is required for communication configuration. Output of RS485 Modbus is connected with serial communication to PC as given in Figure 3. Each process value from the scanner is addressed with Modbus formatting.

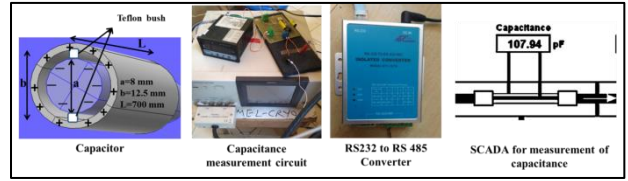

Fig. 3: Capacitance measurement setup

### III. EXPERIMENTAL RESULTS

The designed circuit is developed and simulated, thereafter electronics circuit is tested with several capacitors in the range of 0-600pf and operation frequency of 740 Hz and circuit is calibrated using highly accurate IM 3570 Impedance Analyzer-Hioki L 2000 4 -terminal probe , calibration plot is shown in fig. 4.
The void sensor element is coaxial capacitor and its dimensions are fixed thus the change in dielectric constant gives the change in capacitance which is measured as void fraction for given two phase flow. The medium used in the experiment is nitrogen and its ε varies from $\varepsilon_v = 1$ to $\varepsilon_L = 1.43$
For coaxial cylindrical capacitance probe, the capacitance is given by

$$C = \frac{2\pi L \varepsilon_0 \varepsilon_r}{\ln\left(\frac{b}{a}\right)}$$

Where,
C= Capacitance in pF
L= Length of the capacitor in m
$\varepsilon_r$ = Relative Permittivity of the medium
b= Outer radius in m and
a= inner radius in m
The theoretically estimated value of capacitance is given in Table 2. But, it was found in the experiments that due to additional structural capacitance, the original value deviates from estimated value as shown in table 2.

Table 2: Experimental and theoretical range of capacitance of the void measurement probe

| | Nitrogen | $C_{Theory}$ (pF) | $C_{Experiment}$ (pF) |
|---|---|---|---|
| $\varepsilon_G$ | 1 | 86.79 | 103.59 |
| $\varepsilon_L$ | 1.43 | 124.78 | 148.13 |

In Two-phase Nitrogen dielectric constant (ε) variation with void fraction (α), is measured by,

$$\varepsilon = \varepsilon_v \alpha + \varepsilon_L(1-\alpha)$$

$$\alpha = \frac{(\varepsilon_L - \varepsilon)}{(\varepsilon_L - \varepsilon_v)} \; or \; \alpha = \frac{C_L - C}{C_L - C_v}$$

Where $C_L$ = Capacitance of liquid nitrogen
$C_v$ = Capacitance of vapor nitrogen.
$C$ = Capacitance of two phase mixture.
The values of $C_L$ and $C_v$ are given in table 2. It is observed that void fraction varies linearly with the capacitance ($C$). Figure 5 shows the process flow diagram of the nitrogen two phase flow void fraction measurement experimental setup.

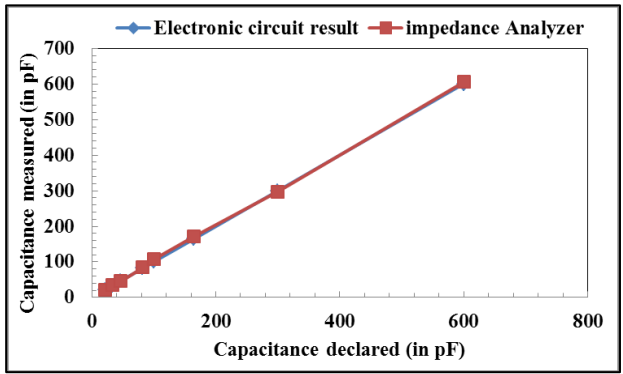

Fig. 4: Comparison of results of our developed electronic circuit and Impedance analyzer

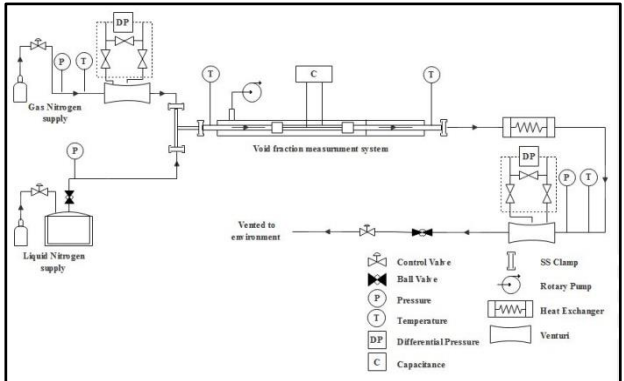

Fig. 5: Experiment setup process flow diagram

## IV. CONCLUSION

An innovative electronic circuit has been designed, developed and tested for its performance while conducting an experiment of the void fraction measurement in two phase nitrogen. Void fraction is measured by measuring the change in the capacitance of coaxial capacitive sensor due to two phase flow. Capacitance of the nitrogen vapor differs from its pure liquid phase due to its dielectric constant. Capacitance variation of the two phase flow in transfer line is from 103pF (room temperature, air medium) to 148.59 pF (pure liquid nitrogen medium). The electronic circuit developed was highly sensitive and accurate with %error of ±0.5%. The measurement range of circuit is 0-600 pF and is calibrated using standard LCR meter. Looking at the accuracy and its precision offered by this developed electronic circuit, it will be useful in applications deal with low capacitance and dynamic capacitance measurement.